\documentclass[twocolumn,superscriptaddress,showkeys,
showpacs,preprintnumbers,amsmath,amssymb,APS]{revtex4}

\usepackage{amssymb,amsmath,amsthm}
\usepackage{epsfig}

\newcommand{\LP}{\left(}
\newcommand{\RP}{\right)}

\newcommand{\ket}[1]{\left|#1\right\rangle}
\newcommand{\om}[2]{\omega_{#1}^{#2}}

\newcommand{\Z}{\mbox{$\mathbb Z$}}

\def\grp#1{\left\langle #1\right\rangle}     

\def\Z{\mathbb{Z}}

\begin{document}
\title{Quantum algorithms for the hidden subgroup problem
on a class of semidirect product groups}

\author{C.M.M. Cosme}%
\email{cmagno@lncc.br}%
\affiliation{Laborat\'{o}rio Nacional de Computa\c{c}\~{a}o
Cient\'{\i}fica (LNCC), C.P. 95113, Petr\'{o}polis, RJ,
25651-075, Brazil}%
\author{R. Portugal}%
\email{portugal@lncc.br}%
\affiliation{Laborat\'{o}rio Nacional de Computa\c{c}\~{a}o
Cient\'{\i}fica (LNCC), C.P. 95113, Petr\'{o}polis, RJ,
25651-075, Brazil}%

\date{\today}

\begin{abstract}
We present efficient quantum algorithms for the hidden subgroup
problem (HSP) on the semidirect product of cyclic groups
$\Z_{p^r}\rtimes_{\phi}\Z_{p^2}$, where $p$ is any odd prime
number and $r$ is any integer such that $r>4$. We also address the
HSP in the group $\Z_{N}\rtimes_{\phi}\Z_{p^2}$, where $N$ is an
integer with a special prime factorization. These quantum
algorithms are exponentially faster than any classical algorithm
for the same purpose.
\end{abstract}

\maketitle


\section{Introduction}

Most of exponentially fast quantum algorithms can be cast into the
hidden subgroup problem (HSP), which is considered a paradigm for
the development of quantum algorithms. The HSP on the finite group
$G$ can be described as follows. Let $X$ be a finite set and $f:G
\rightarrow X$ a function such that $f(g_1)=f(g_2)$ if and only if
$g_1$ and $g_2$ are in the same left coset of some subgroup $H$ in
$G$. The problem consists in determining generators for $H$ by
querying the function $f$. If the computational complexity of the
algorithm is $O(\log |G|)$ considering that each query counts as
one computational step, we say that the HSP is solved efficiently.
It is usual to say that the function $f$ hides the subgroup $H$ in
$G$. Simon~\cite{Simon} and Shor~\cite{Shor} algorithms solve
special cases of the abelian HSP. When the group $G$ is abelian,
there is an efficient solution in the general
case~\cite{Kitaev,Lomont}.

If $G$ is not abelian but is ``close'' to abelian in some sense,
the HSP may be solved by reduction to the abelian HSP. There are
many examples of this method in the
literature~\cite{Ivanyos1,Ivanyos2}, which may not employ new
quantum algorithms directly. Recently, Inui and Le
Gall~\cite{Inui} have presented an efficient algorithm for the HSP
on the group $\Z_{p^r}\rtimes_{\phi}\Z_{p}$ for odd prime $p$ and
positive integer $r$ employing direct quantum techniques. Using
the classification of the subgroups of
$\Z_{p^r}\rtimes_{\phi}\Z_{p}$ presented in~\cite{Inui}, it is
possible to use the Ettinger-H{\o}yer reduction~\cite{Ettinger} to
simplify the algorithm to the abelian HSP~\cite{Bacon}. There is
an alternative form to reduce Inui and Le Gall's algorithm to the
abelian HSP by employing the results of Ref.~\cite{Ivanyos1} on
normal subgroups of solvable groups. We discuss this method in
Sec.~\ref{sec5}. In Ref.~\cite{Chi}, the authors have extended the
solution of the HSP to the group $\Z_{N}\rtimes_{\phi}\Z_{p}$
where $N$ is factorized as $N=p_1^{r_1}\cdots p_n^{r_n}$ and $p$
does not divide each $p_i-1$ for $1\leq i\leq n$.

In the present work we address the HSP on the group
$\Z_{p^r}\rtimes_{\phi}\Z_{p^2}$ for any odd prime $p$ and any
integer $r$ such that $r>4$. We present an efficient solution for
the HSP both by using direct quantum algorithms and by reducing to
the abelian HSP. We also address the HSP on the group
$\Z_{N}\rtimes_{\phi}\Z_{p^2}$ where $N$ is factorized as
$N=p_1^{r_1}\cdots p_n^{r_n}$ and $p$ does not divide each
$p_i-1$.

This paper is organized as follows. In Sec.~\ref{sec1} we describe
the structure of the group $\Z_{p^r}\rtimes_{\phi}\Z_{p^2}$ and
give the complete list of subgroups. There are two classes of
non-trivial isomorphic groups depending on the homomorphism
$\phi$. In Sec.~\ref{sec2} we list the abelian subgroups that can
be used to run the abelian HSP in order to obtain information
about $H$ (the hidden subgroup). In Sec.~\ref{sec3} we address the
case when $H$ is cyclic in the first class of groups. In
Sec.~\ref{sec4} we address the non-cyclic case also in the first
class. In Sec.~\ref{sec5} we address the second class. In
Sec.~\ref{sec5dot5} we address the HSP on the group
$\Z_{N}\rtimes_{\phi}\Z_{p^2}$ with some restrictions over the
values of the integer $N$. In Sec.~\ref{sec6} we present our
conclusions.

\section{The structure of $\Z_{p^r}\rtimes_{\phi}\Z_{p^2}$}\label{sec1}

The group $\Z_{p^r}\rtimes_{\phi}\Z_{p^2}$ is the set  $\{(a,b), a
\in \Z_{p^r}, b\in \Z_{p^2}\}$ with the group operation
$(a_{1},b_{1})(a_{2},b_{2})=(a_{1}+\phi(b_{1})(a_{2}),~b_{1}+b_{2})$,
where $\phi$ is any group homomorphism from $\Z_{p^2}$ into the
group of automorphims of $\Z_{p^r}$. The group
$\Z_{p^r}\rtimes_{\phi}\Z_{p^2}$ is generated by $x=(1,0)$ and
$y=(0,1)$ and $\phi$ is completely determined by the value
$\alpha=\phi(1)(1)$, which is in $\Z_{p}^{\ast}$. For $r>4$ the
possible values of $\alpha$ are $\tau p^{r-2}+1$, where
$0\leq\tau<p^2$. There are two classes of non-trivial isomorphic
groups.

Class (1) is characterized by the values of $\tau$ such that
gcd$(\tau,p^2)=1$. We take $\tau=1$, that is $\alpha=p^{r-2}+1$
with no loss of generality. Using that
$(a,b)=x^a y^b$ and $\phi(b)(a)=a\alpha^b$ we get %
$$ \LP x^{a_1}y^{b_1} \RP\LP x^{a_2}y^{b_2} \RP = x^{a_1+a_2(b_1
p^{r-2}+1)}y^{b_1+b_2}. $$

Class (2) is characterized by the values of $\tau$ such that
gcd$(\tau,p^2)=p$. We take $\tau=p$, that is, $\alpha=p^{r-1}+1$
with no loss of generality. The product of two group elements is
given by %
$$ \LP x^{a_1}y^{b_1} \RP\LP x^{a_2}y^{b_2} \RP = x^{a_1+a_2(b_1
p^{r-1}+1)}y^{b_1+b_2}. $$

The subgroups of $\Z_{p^r}\rtimes_{\phi}\Z_{p^2}$ for both classes
have the following forms:
\begin{equation}\grp{x^{p^i}}\,\,{\rm or}\,\,\grp{x^{tp^i}y^{p^j}},\label{sg1}\end{equation}%
where $0\leq i\leq r$, $0\leq j \leq 1$, and $t\in\Z_{p^l}^{\ast}$
where $l=\min\{r-i,\ 2-j\}$,
\begin{equation}\grp{x^{p^i},y^{p^j}},\label{sg2}\end{equation}%
where $0\leq i< r$, $0\leq j\leq 1$, and
\begin{equation}\grp{x^{tp^i}y,x^{p^{i+1}}},\label{sg3}\end{equation}%
where $0\leq i< r$ and $t\in\Z_{p}^\ast$.

The subgroups are characterized by 3 parameters: $t$, $i$, and
$j$. This is too complex for any direct attempt to build a quantum
algorithm. As usual, the strategy we will use is the following.
Let $G$ be an abelian subgroup of $\Z_{p^r}\rtimes_{\phi}\Z_{p^2}$
and suppose that the function $f$ hides $H$ in
$\Z_{p^r}\rtimes_{\phi}\Z_{p^2}$. If $H$ is subgroup of $G$, then
the generators of $H$ can be found efficiently by restricting $f$
to $G$ and by employing the algorithms of the abelian HSP. If $H$
is not a subgroup of $G$, this method finds efficiently the
generators of $H\cap G$, which may yield partial information about
$H$. We will see that this strategy eliminates the parameter $i$.
The remaining parameters $t$ and $j$ will be determined by two
methods: (1) direct quantum algorithms and (2) reduction using the
algorithms described in Ref.~\cite{Ivanyos1}.

\section{Reduction to abelian subgroups}\label{sec2}

The abelian subgroups of $\Z_{p^r}\rtimes_{\phi}\Z_{p^2}$ that we
use in this part of the reduction are $\grp{x}$ and $\grp{y}$. Let
$H_x=H\cap\grp{x}$ and $H_y=H\cap\grp{y}$. Function $f_x$ defined
by $f_x(a)=f(a,0)$ hides $H_x$ in $\Z_{p^r}$. Function $f_y$
defined by $f_y(b)=f(0,b)$ hides $H_y$ in $\Z_{p^2}$. The solution
of the abelian HSP on $\Z_{p^r}$ and $\Z_{p^2}$ with oracles $f_x$
and $f_y$ respectively determines the generators for $H_x$ and
$H_y$. The form of these groups are $H_x=\grp{x^{p^m}}$ and
$H_y=\grp{y^{p^n}}$, where $0\leq m\leq r$ e $0\leq n\leq 2$.
Therefore, the abelian reduction yields the values of $m$ and $n$
efficiently.

Using the values of $m$ and $n$ it is possible to determine
whether $H$ is cyclic or not. If $H$ is generated by two elements,
either formula~(\ref{sg2}) or (\ref{sg3}), one can verify that
$H_x$ and $H_y$ cannot be the trivial group. One eventually
concludes that if $m=r$ or $n=2$ then $H$ is cyclic, that is, $H$
is described by formula~(\ref{sg1}); and if $0\le m<r$ and $0\le
n<2$ then $H$ is not cyclic, that is, either $H$ is described by
formula~(\ref{sg2}) or by formula~(\ref{sg3}).

In the following sections we address the groups in class (1). In
Sec.~\ref{sec5} we address the groups in class (2).

\section{The cyclic case}\label{sec3}

Suppose that $m=r$ or $n=2$. By running the abelian HSP on the
subgroup $\grp{x^{p^2},y}$ with the oracle $f$ restricted to this
subgroup, either we determine generators for $H$ or in the worst
case we obtain partial information about $H$. The way to proceed
at this point is to calculate the intersection of the groups of
the form given by formula~(\ref{sg1}) with $\grp{x^{p^2},y}$. We
split into 3 cases: (\textit{i}) $m=r$ and $n=2$, (\textit{ii})
$m=r$ and $n<2$, and (\textit{iii}) $m<r$ and $n=2$.

In case (\textit{i}), the groups of the form given by
formula~(\ref{sg1}) are subgroups of $\grp{x^{p^2},y}$, therefore
the abelian reduction provides the values of the parameters $t$,
$i$, and $j$. The possible forms of $H$ are $\grp{x^{t p^{r-2}}
y}$ where $t\in\Z_{p^2}^\ast$, $\grp{x^{t p^{r-1}} y^p}$ where
$t\in\Z_{p}^\ast$, and $\grp{\,(0,0)\,}$.

Case (\textit{ii}) is similar to case (\textit{i}), because $H$ is
a subgroup of $\grp{x^{p^2},y}$. The abelian reduction provides
the values of the parameters $t$, $i$, and $j$. The possible forms
of $H$ are $\grp{y}$ when $n=0$ and $\grp{x^{t p^{r+j-1}}
y^{p^j}}$ when $n=1$ where $t\in\Z_{p^{2-j}}^\ast$, $0\leq j\leq
1$.

In case (\textit{iii}), if $m=0$ or $m\ge 4$, $H$ is subgroup of
$\grp{x^{p^2},y}$. The abelian reduction provides completely the
values of the parameters $t$, $i$, and $j$. The possible forms of
$H$ are $\grp{x}$ when $m=0$ and $\grp{x^{p^m}}$ or
$\grp{x^{tp^{m-2+j}}y^{p^j}}$ when $m\ge 4$ where
$t\in\Z_{p^{2-j}}^\ast$ and $0\leq j\leq 1$. The remaining cases
are $m=1$, $m=2$, and $m=3$.

If $m=1$, the abelian reduction eliminates the parameter $i$. The
values of $t$ and $j$ remain unknown. The possible forms of $H$
are $\grp{x^p}$ and $\grp{x^ty^p}$ where $t\in\Z_{p}^\ast$. We
have to decide between these two forms and in the last form we
have to determine the value of $t$. We proceed by employing a
quantum algorithm. We prepare the quantum computer in the initial
state
\begin{equation}
\ket{\psi_1}=\frac{1}{\sqrt{p^{3}}}\sum_{a'=0}^{p-1}
                \sum_{b'=0}^{p^2-1}\ket{a'}\ket{b'}\ket{f(x^{a'}y^{b'})}.
\end{equation}
Now we measure the third register in the computational basis. The
result depends on the form of $H$. Suppose by now that
$H=\grp{x^ty^p}$, for some $t\in\Z_{p}^\ast$. The result is
\begin{equation}
\ket{\psi_2}=\frac{1}{\sqrt{p}}\sum_{l=0}^{p-1}\ket{(a_0+t\,l)\,{\rm
mod}\,p}\ket{b_0+p\,l},
\end{equation}
for some $a_0$ and $b_0$ such that $0\le a_0< p$ and $0\le b_0<
p^2$ randomly distributed. We have disregarded the third register
since it will be irrelevant from now on. Now we apply the Fourier
transform operator F$_p\otimes$F$_{p^2}$ on state $\ket{\psi_2}$.
The result is
\begin{equation}
\ket{\psi_3}=\frac{1}{\sqrt{p^3}}\sum_{a=0}^{p-1}\sum_{\begin{subarray}{c}{b=0}\\{
at+b\equiv 0\,{\rm mod}\,
p}\end{subarray}}^{p^2-1}\om{p}{aa_0}\om{p^2}{bb_0}
    \ket{a}\ket{b},
\end{equation}
where $\omega_p$ ($\omega_{p^2}$) is the primitive $p$-root
($p^2$-root) of the unity. Now we measure the state $\ket{\psi_3}$
in the computational basis and get values $a$ and $b$ such that
$at+b\equiv 0\,{\rm mod}\,p$. If $a\not\equiv 0\,{\rm mod}\,p$,
then we calculate $t_1=-a^{(-1)}b\,{\rm mod}\,p$. If $f(x^{t_1}
y^p)=f(1)$ then $H=\grp{x^{t_1}y^p}$ otherwise $H=\grp{x^p}$. The
success probability is $1-1/p$.

If $m=2$ the possible forms of $H$ are $\grp{x^{p^2}}$, $\grp{x^{t
p} y^p}$ where $t\in\Z_{p}^\ast$, and $\grp{x^{t} y}$ where
$t\in\Z_{p^{2}}^\ast$. We have to decide among those forms and
then we have to determine the value of $t$. We proceed by using a
quantum algorithm. We use the initial state
\begin{equation}
\ket{\psi_1}=\frac{1}{p^2}\sum_{a'=0}^{p^2-1}
                \sum_{b'=0}^{p^2-1}\ket{a'}\ket{b'}\ket{f(x^{a'}y^{b'})}
\end{equation}
and after measuring the third register we apply the Fourier
transform operator F$_{p^2}\otimes$F$_{p^2}$. The way to proceed
is similar to the case $m=1$. At the end, if $a\not\equiv 0\,{\rm
mod}\,p$, then we calculate $t_1=-a^{(-1)}b\,{\rm mod}\,p^2$ and
$t_2=-a^{(-1)}b\,{\rm mod}\,p$. If $f(x^{t_1} y)=f(1)$ then
$H=\grp{x^{t_1}y}$. If $f(x^{t_2 p} y^p)=f(1)$ then $H=\grp{x^{t_2
p}y^p}$ otherwise $H=\grp{x^{p^2}}$.

If $m=3$ the possible forms of $H$ are $\grp{x^{p^3}}$, $\grp{x^{t
p^2} y^p}$ where $t\in\Z_{p}^\ast$, and $\grp{x^{t p} y}$ where
$t\in\Z_{p^{2}}^\ast$. This case is very similar to the case
$m=2$. The only differences are that the first sum of the state
$\ket{\psi_1}$ runs from 0 to $p^3-1$ and the Fourier transform
operator is F$_{p^3}\otimes$F$_{p^2}$. The remaining analysis of
the algorithm is similar to the case $m=2$.

\section{The non-cyclic case}\label{sec4}

Suppose that $0\le m< r$ and $0\le n<2$.  Either $H$ has the form
given by formula~(\ref{sg2}) or the form of formula~(\ref{sg3}).
If $n=0$ then there is only one possible form which is
$H=\grp{x^{p^m},y}$. From now on we consider the case $n=1$. If
$m=0$ then there is only one possibility which is $H=\grp{x,y^p}$.
For $m\ge 3$, $H$ is a subgroup of $\grp{x^{p^2},y}$, therefore
the abelian reduction over $\grp{x^{p^2},y}$ provides the values
of the parameters $t$, $i$, and $j$. The possible forms of $H$ are
$\grp{x^{p^m}, y^p}$ and $H=\grp{x^{tp^{m-1}}y,x^{p^m}}$. The
remaining cases are $m=1$ and $m=2$.

If $m=1$ then either $H=\grp{x^{p},y^{p}}$ or
$H=\grp{x^{t}y,x^{p}}$, $t\in\Z_{p}^\ast$. The abelian reduction
does not determine between those two forms and does not provide
the value of $t$. We proceed by employing a quantum algorithm. We
prepare the quantum computer in the following initial state
\begin{equation}
\ket{\psi_1}=\frac{1}{{p}}\sum_{a'=0}^{p-1}
                \sum_{b'=0}^{p-1}\ket{a'}\ket{b'}\ket{f(x^{a'}y^{b'})}.
\end{equation}
We measure the third register in the computational basis. The result
depends on the form of $H$. Suppose by now that
$H=\grp{x^{t}y,x^{p}}$, for some $t\in\Z_{p}^\ast$. The result is
\begin{equation}
\ket{\psi_2}=\frac{1}{\sqrt{p}}\sum_{l=0}^{p-1}\ket{(a_0+t\,l)\,{\rm
mod}\,p}\ket{l},
\end{equation}
for some $a_0$ such that $0\le a_0< p$ randomly distributed. Now
we apply the Fourier transform operator F$_p\otimes$F$_{p}$ on
state $\ket{\psi_2}$. The result is
\begin{equation}
    \ket{\psi_3}=\frac{1}{p}\sum_{\begin{subarray}{c}{a,b=0}\\{
at+b\equiv 0\,{\rm
mod}\,p}\end{subarray}}^{p-1}\om{p}{aa_0}\ket{a}\ket{b}.
\end{equation}
Now we measure the state $\ket{\psi_3}$ in the computational basis
and get values $a$ and $b$ such that $at+b\equiv 0\,{\rm mod}\,p$.
If $a\not\equiv 0\,{\rm mod}\,p$, then we calculate
$t_1=-a^{(-1)}b\,{\rm mod}\,p$. If $f(x^{t_1} y)=f(1)$ then
$H=\grp{x^{t_1}y,x^{p}}$ otherwise $H=\grp{x^{p},y^{p}}$. The
success probability is $1-1/p$.

If $m=2$ then either $H=\grp{x^{p^2},y^{p}}$ or
$H=\grp{x^{tp}y,x^{p^2}}$ where $t\in\Z_{p}^\ast$. For this case
we use the same strategy of the case $m=1$ with some minor
differences.

The use of direct quantum algorithms can be avoided if one notes
that all subgroups for which we have employed quantum algorithms
in Secs.~\ref{sec3} and \ref{sec4} are normal in
$\Z_{p^r}\rtimes_{\phi}\Z_{p^2}$. The list of those subgroups is
$\grp{x^{p^k}}$ $1\leq k\leq 3$, $\grp{x^p,y^p}$,
$\grp{x^{p^2},y^p}$, $\grp{x^{t} y^p}$, $\grp{x^{t p} y^p}$,
$\grp{x^{t p^2} y^p}$, $\grp{x^t y,x^p}$, $\grp{x^{tp}
y,x^{p^2}}$, $t\in\Z_{p}^\ast$; $\grp{x^{t} y}$, $\grp{x^{t p}
y}$, $t\in\Z_{p^{2}}^\ast$. It is easy to verify because they
contain the commutator group of $\Z_{p^r}\rtimes_{\phi}\Z_{p^2}$
which is $\grp{x^{p^{r-2}}}$. We know that
$\Z_{p^r}\rtimes_{\phi}\Z_{p^2}$ is a finite $p$-group, therefore
it is solvable. Theorem 7 of Ref.~\cite{Ivanyos1} states that if
$H$ is a normal hidden subgroup of a solvable group $G$, then the
generators of $H$ can be found by a quantum algorithm in time
polynomial in $\log |G|$. In the proof of their result, they show
constructively how the problem reduces to the abelian HSP. From
this argument we conclude that the HSP on
$\Z_{p^r}\rtimes_{\phi}\Z_{p^2}$ for odd prime $p$ and $r>4$ can
be fully reduced to the abelian HSP.

\section{The groups in class (2)}\label{sec5}

The groups in class (2) are somewhat simpler than the ones in
class (1). The main reason is that the subgroup $\grp{y^p}$ is
normal in $\Z_{p^r}\rtimes_{\phi}\Z_{p^2}$ and the quotient group
$(\Z_{p^r}\rtimes_{\phi}\Z_{p^2})/\grp{y^p}$ is isomorphic to
$\Z_{p^r}\rtimes\Z_{p}$ when we take the homomorphism $\phi$ that
characterizes class (2).

Let us show how the HSP on the groups in class (2) reduces to the
abelian HSP. In Sec.~\ref{sec2} we have showed how to determine
the values of parameters $m$ and $n$. Using these values, we can
know in advance whether (\textit{i}) $H$ is completely determined,
(\textit{ii}) $H$ is a subgroup of $\grp{x^p,y}$, or
(\textit{iii}) $H$ is normal. Case (\textit{i}) occurs when $m$ or
$n$ is zero. If $m=0$ then $H=\grp{x,y^{p^n}}$ where $0\leq n\leq
2$. If $n=0$, then $H=\grp{x^{p^m},y}$ where $0 < m\leq r$.

Case (\textit{ii}) occurs in the following cases. If $m=r$ and
$0\leq n\leq 2$ then $H=\grp{x^{tp^{r-n+j}}y^{p^j}}$ where either
$t=1$ when $j=n$ or $t\in\Z_{p^{n-j}}^\ast$ when $0\leq j<n$. If
$3\le m<r$ and $n=2$ then $H=\grp{x^{tp^{m+j-2}}y^{p^j}}$ where
either $t=1$ when $j=2$ or $t\in\Z_{p^{2-j}}^\ast$ when $0\leq
j<2$. If $2\le m<r$ and $n=1$ then either $H=\grp{x^{p^{m}},y^p}$
or $H=\grp{x^{tp^{m-1}}y,x^{p^m}}$ where $t\in\Z_{p}^\ast$.

Case (\textit{iii}) occurs in the following cases. If $1 \leq m
\leq 2$ and $n=2$ then $H=\grp{x^{tp^{m+j-2}}y^{p^j}}$ where
$0\leq j \leq 2$. If $m=n=1$ then either $H=\grp{x^p, y^p}$ or
$H=\grp{x^t y,x^p}$.

The strategy to solve the HSP on $\Z_{p^r}\rtimes_{\phi}\Z_{p^2}$
in class (2) is the following. In case (\textit{i}) we are done.
In case (\textit{ii}) we run the abelian HSP with the function $f$
restricted to the subgroup $\grp{x^p,y}$. This procedure
determines completely the parameters of the generators of $H$. In
case (\textit{iii}) we employ the algorithms described in
Ref.~\cite{Ivanyos1}. We can use them because we know in advance
that $H$ is normal.

The same strategy works for the group $\Z_{p^r}\rtimes\Z_{p}$
which was addressed in Ref.~\cite{Inui}. This group has the
following properties. All proper subgroups are abelian and the
maximal subgroups are normal. By running the abelian HSP on the
subgroups $\grp{x}$, $\grp{y}$, $\grp{x^p,y}$, either one obtains
the generators of $H$ or learns that $H$ is normal. In the latter
case one employs the methods of Ref.~\cite{Ivanyos1}.

\section{Groups of the form $\Z_{N}\rtimes_{\phi}\Z_{p^2}$}\label{sec5dot5}

The same kind of reduction presented in Ref.~\cite{Chi} applies to
the group $\Z_{N}\rtimes{_\phi}\Z_{p^2}$, where the prime
factorization of $N$ is $p_1^{r_1}\cdots p_n^{r_n}$ and $p$ does
not divide each $p_j-1$, $1\leq j \leq n$.

The following result holds is this case. If $p$ and $q$ are
distinct primes satisfying $p\nmid (q-1)$ then
$(\Z_{q^s}\times\Z_{p^r})\rtimes_{\phi}\Z_{p^2}$ is isomorphic to
$\Z_{q^s}\times(\Z_{p^r}\rtimes_{\psi}\Z_{p^2})$ for some
homomorphism $\psi$ from $\Z_{p^2}$ into the group of automorphims
of $\Z_{p^r}$. The proof is similar to the one presented in Lemma
2 of Ref.~\cite{Chi}. Because $p$ must divide the order of
$\Z_{N}^{\ast}$, we can choose $p_1=p$ with no loss of generality.
Using those results, it is straightforward to show that %
$$\Z_{N}\rtimes_{\phi}\Z_{p^2} \cong \Z_{p_2^{r_2}}\times \cdots
\times \Z_{p_n^{r_n}}\times
(\Z_{p_1^{r_1}}\rtimes_{\psi}\Z_{p^2}).$$ %
The orders of the groups in the direct product of the above
isomorphism are relatively prime. Therefore, the HSP on
$\Z_{N}\rtimes_{\phi}\Z_{p^2}$ reduces to the HSP on each factor.
Either the factor is an abelian group or it is the group
$\Z_{p_1^{r_1}}\rtimes_{\psi}\Z_{p^2}$, which was addressed in
this paper if $r_1>4$.

\

\section{Conclusions}\label{sec6}

We have described efficient quantum algorithms for the HSP on the
group $\Z_{p^r}\rtimes_{\phi}\Z_{p^2}$, where $p$ is any odd prime
number and $r$ is any integer such that $r>4$. The method relies
on the classification of all subgroups of
$\Z_{p^r}\rtimes_{\phi}\Z_{p^2}$. The subgroups are characterized
by three parameters. By using reductions to the abelian HSP, the
number of independent parameters decreases and the values of the
remaining ones are found either by employing direct quantum
algorithms or by using the reduction described in
Ref.~\cite{Ivanyos1}. We have also addressed the HSP on the group
$\Z_{N}\rtimes_{\phi}\Z_{p^2}$ where $N$ is factorized as
$N=p_1^{r_1}\cdots p_n^{r_n}$ and $p$ does not divide each
$p_i-1$, by employing an isomorphism between
$\Z_{N}\rtimes_{\phi}\Z_{p^2}$ and the direct product of
$\Z_{p^r}\rtimes_{\phi}\Z_{p^2}$ with cyclic groups.

The computational complexity of the algorithm can be bounded by
the following analysis. The order of the group
$\Z_{p^r}\rtimes_{\phi}\Z_{p^2}$ is $p^{(r+2)}$. We have employed
abelian reductions, direct quantum algorithms, and the reduction
described in Ref.~\cite{Ivanyos1}. In all those parts we can
guarantee that the complexity is $O$(poly($(r+2)\log p$)).
Therefore the overall complexity of the algorithm for solving the
HSP on $\Z_{p^r}\rtimes_{\phi}\Z_{p^2}$ is $O$(poly($(r+2)\log
p$)). The algorithm is probabilistic and we guarantee a success
probability greater than $1/2$.

We are currently addressing the HSP on the group
$\Z_{p^r}\rtimes\Z_{p^s}$ for any odd prime $p$ and integers $r$
and $s$ such that $r>2 s$. This case seems to be a straightforward
generalization of the algorithms presented in this work.

\section*{Acknowledgments}
We thank Guilherme Leal and Demerson N. Gon\c{c}alves for useful
discussions. This work was funded by FAPERJ and CNPq.

\end{document}